\documentstyle[12pt,epsf]{article}

\begin{document}
 
\begin{center}

{\LARGE Flow equations in light-front QCD\\}

\vskip 20pt

{\Large Elena Gubankova \\}
\bigskip

{\it Department of Physics, North Carolina State University, Raleigh,
     NC 27696-8202 } 

\bigskip
\end{center}

\date{\today}

\vspace{2cm}

\begin{abstract}

Light-front QCD is studied by the method of flow equations.
Dynamical gluon mass is generated, which evolves with the cut-off
according to renormalization group equation. 
Eliminating by flow equations the quark gluon coupling 
with the dynamical gluon mode, one obtains 
an effective interaction between quark and antiquark
which exibits the Coulomb and confining singularities.
The scale, which regulates the light-front IR singularities
in the gluon sector, defines the string tension
of confining interaction. The mechanism of confinement 
in the light-front formalism is suggested, based 
on the singular nature of the light-front gauge.

\end{abstract}

\newpage
\section{Introduction}
\label{sec:intr}

Quantum Cromodynamics (QCD) is a widely accepted theory of strong interactions.
This wide-spread acceptance is based on the success of Feynman rules 
of perturbative covariant calculations, which provided convincing agreement
between perturbative QCD and experiment. However there is a gap between the perturbative
behavior of QCD and its low-energy limit, where perturbation theory breaks down
and the physical observables such as mass spectrum and decay width are predicted based
on phenomenology. In the strong coupling regime non-perturbative methods
are required. The non-perturbative solution of a bound state problem can be obtained
directly only by using the Hamiltonian formalism. 

It is crucial to a successful non-perturbative solution that it exposes the three important
long range properties of QCD: confinement, spontaneous chiral symmetry breaking,
and the topological structure. We show in this work, that the non-perturbative method
of flow equations, when applied to the QCD Hamiltonian in the light-front quantization,
provides some understanding of the physics of confinement in the Scr{\"o}dinger picture.

There were several successful attempts to reveal the mechanism of confinement
in the Schr{\"o}dinger picture, using a special ansatz for a vacuum wave functional
and integrating over all possible gauge configurations \cite{Diakonov}. 
We believe that the light-front quantization provides an alternative formalism,
where it is possible to isolate the degrees of freedom that are responsible
for the long-range properties of QCD. In the light-front quantized QCD the topological structure
is carried by the zero mode of $A^+$ \cite{Pinsky}, 
instead of (nontrivial) gluon vacuum configurations as in other gauges.

The confinement mechanism in the light-front QCD was suggested several years ago
\cite{BrisudovaPerry}, based on the fact that $QCD_{3+1}$ already has a confining interaction
term in the light-front Hamiltonian, the instantaneous four Fermion interaction,
which is the confining interaction in $QCD_{1+1}$. The authors argue, that in QCD
the second order quark glue interaction, which appears through similarity renormalization,
does not cancel the instantaneous interaction as it does in perturbation theory.
The singular part of the uncanceled instantaneous interaction $(\sim 1/q^{+ 2})$,
produces a logarithmic potential of the form
\begin{eqnarray}
V(\vec{r})\sim \frac{2\omega a(\hat{e}_r)}{\pi}\log r
\end{eqnarray} 
where $a$ is equal $1$ for the radial tensor $\hat{e}_r$ along the $z$-axis
and it equals $2$ when $\vec{r}$ is purely transverse;
$\omega$ has dimension of energy and sets up a scale for the potential.
This potential is confining. It is boost invariant, but it is not rotational symmetric
that is confusing. It was also pointed in \cite{Pinsky}, that the instantaneous interaction
of the off diagonal currents is modified by the topological properties of the theory,
and confinement is destroyed for these currents in $QCD_{1+1}$ and also in $QCD_{3+1}$.

The basic idea of our calculations is to use the 'singular' nature of the light-front gauge
in conjunction with non-perturbative renormalization provided by flow equations.
The main conceptual complication to study renormalization group (RG) flow 
of the effective QCD Hamiltonian is the lack of a well defined initial condition.
We study the RG equation for an effective gluon mass, given some parameter mass $\tilde{\mu}$
for the gluon mass in renormalization point. This mass $\tilde{\mu}$ is used as a parameter
and is taken to zero at the end of calculations. Apart from the known perturbative
gluon correction in the second order, we obtain also the term which is a function 
of the ultraviolet cut-off $\lambda$ and the mass parameter $\tilde{\mu}$. We associate
the second term with 'non-perturbative' gluon mass correction.
In the light-front quantization the gluon non-abelian gauge interactions produce
severe infrared divergences in the effective gluon mass. The subtle point is, that
we use an additional scale $u$, $u\ll\lambda$, as suggested by Zhang and Harindranath
in \cite{ZhangHarindranath}, to regulate these divergences. One ends up 
with an effective 'non-perturbative' gluon mass, which is the function of the parameters
$u$ and $\tilde{\mu}$ and the cut-off $\lambda$, $\mu^2_{NP}(\lambda;u,\tilde{\mu})$.
In the limit $\tilde{\mu}\rightarrow 0$, only the non-abelian part of the effective gluon mass,
regulated by the scale $u$, survives. This is the crucial difference between QED and QCD.
In QED the perturbative gluon mass correction can be removed by perturbative renormalization,
i.e. renormalized photon stayes massless. In QCD by absorbing the leading cut-off dependence
in the second order mass counterterm, we are still left with non-perturbative mass correction.
It turns out that this 'dressing'  of an effective gluon playes an important role
in the effective interaction between quarks.

We simulate an effective interaction in QCD between probe static quark and antiquark
as an exchange of 'non-perturbative' gluon (gluon flux) with a nonzero 
effective mass $\mu^2_{NP}(\lambda;u,\tilde{\mu})$ (effective energy),
which evolves with the cut-off $\lambda$ according to RG equation. Eliminating
by flow equations the quark gluon coupling with an effective gluon mass,
one obtains the quark-antiquark potential which includes two pieces:
the short-range part describes the perturbative one-gluon exchange and is analog
of the perturbative interaction in QED \cite{GubankovaWegner};
the long-range part arises due to the non-perturbative gluon 'dressing', i.e.  
due to the dependence of gluon effective mass on the cut-off.
For the vanishing mass parameter, $\tilde{\mu}\rightarrow 0$,
an effective $q\bar{q}$-potential is given by 
a sum of Coulomb and linear rising confining interactions
\begin{eqnarray}
V(\vec{r})= -C_f\frac{\alpha_s}{r} +\sigma\cdot r
\end{eqnarray}
The parameter $u$ sets up a scale for the long-range part of interaction:
it defines the string tension of the confining term, $\sigma\sim u^2$.
Though the calculations are performed in the light-front frame,
the resulting effective interaction manifests rotational symmetry.

The article is organized in the following way: the first section
introduces the problem and sets up a scheme;
in the second section a gluon gap equation is obtained
and solved for an effective gluon mass;
we obtain an effective potential between quark and antiquark
in the third section.

\section{Flow equations in QCD}
\label{sec:1}
 
Flow equations were discussed in great detail
in application to QED in the previous work \cite{GubankovaWegner}.
Here we point out the differnce between QED and QCD.
For simplicity we consider only the abelian part of QCD Hamiltonian.
Flow equations for the Hamiltonian $H=H_d+(H-H_d)$ read \cite{Wegner}
\begin{eqnarray}
&& \frac{dH(l)}{dl} = [\eta(l), H(l)]
\nonumber\\
&& \eta(l)= [H_d(l), H(l)] 
\,,\label{eq:1.1}\end{eqnarray}  
where $eta$ is the generator of unitary transformation,
which eliminates the particle number changing part of the Hamiltonian,
$(H-H_d)$; the $H_d$ includes all particle number conserving terms;
$l$ is the flow parameter, which changes from $l=0$ corresponding to
the initial canonical Hamiltonian to $l\rightarrow\infty$
with block-diagonal Hamiltonian $H(l\rightarrow\infty)=H_d(l\rightarrow\infty)$.

It is always possible to divide the complete Fock space 
(particle number space) into two arbitrary subspaces,
$P$ and $Q$ space. The Hamiltonian matrix reads
\begin{eqnarray}
   H = \left(
   \begin{array}{cc}
      PHP & PHQ \\
      QHP & QHQ
   \end{array}\right)
\,,\label{eq:1.2}\end{eqnarray}
where $P$ and $Q=1-P$ are projection operators.  
For the (abelian) QCD the content of sectors is given 
\begin{eqnarray}
P|\psi\rangle &=& |g\rangle
\nonumber\\
Q|\psi\rangle &=& |q\bar{q}\rangle
\,,\label{eq:1.3}\end{eqnarray}  
with simbols $g$ and $q$ standing for gluon and quark, respectively.
Therefore the matrix elements of $PHQ$ descibe quark gluon coupling,
$PHP$ stands for gluon effective energy, and $QHQ$ descibes
$q\bar{q}$ effective interaction.
When the Hamiltonian matrix is subject to the unitary transformation Eq. (\ref{eq:1.1}),
the secor Hamiltonians become the functions of the flow parameter $l$.

Suppose we know approximately the eigenstates of the sector Hamiltonians
$PH(l)P$ and $QH(l)Q$ and their eigenvalues $E_p(l)$ and $E_q(l)$.
The indeces $p$ and $q$ run over all states in the $P$ and $Q$ space,
respectively. Suppose further, that this basis is $l$-independent, i.e.
we assume, that the off-diagonal matrix elements $h_{pp'}$ and $h_{qq'}$
of $PHP$ and $QHQ$ are small.
For the particle number conserving sector we keep all the terms in flow equation,
while for the particle number changing sector we neglect the small
off-diagonal matrix elements $h_{pp'}$ and $h_{qq'}$ and take into account
only the diagonal matrix elements $E_p$ and $E_{q}$ on the right-hand side
of flow equation. Flow equations for the matrix elements of 
the particle number conserving and particle number changing sectors 
read, respectively,
\begin{eqnarray}
\frac{dh_{pp'}(l)}{dl} &=& \sum_{q}\left(
\eta_{pq}(l)h_{qp'}(l)-h_{pq}(l)\eta_{qp'}(l)
\right)
\nonumber\\
\frac{dh_{pq}(l)}{dl} &=& -\left(
E_p(l)-E_q(l) \right) \eta_{pq}(l)
\,,\label{eq:1.4}\end{eqnarray}  
and the analagous equation for $h_{qq'}$.
Here the generator is chosen in a more general, than Eq. (\ref{eq:1.1}), form
\begin{eqnarray}
\eta_{pq}(l) &=& -\frac{h_{pq}(l)}{E_p(l)-E_q(l)}\frac{d}{dl}
\left( \ln f(z_{pq}(l)) \right)
\nonumber\\
z_{pq}(l) &=& l\left( E_p(l)-E_q(l) \right)^2
\,,\label{eq:1.5}\end{eqnarray}  
where $f(z)$ is the similarity function with the properties
\begin{eqnarray}
&& f(0) =1
\nonumber\\
&& f(z\rightarrow\infty) =0
\,.\label{eq:1.6}\end{eqnarray}
We take into account in the similarity factor the dependence
of the energies on the flow parameter. This is the crucial difference
between QED and QCD.
We show in the next section that $E_p(l)$ playes the role of the effective energy
(effective mass) of gluon.
In QED this dependence can be removed by perturbative renormalization,
so that one works in terms of renormalized energy (mass) operators 
which are fitted to the physical values. In QCD only the perturbative
energy (mass) correction can be absorbed by the counterterm. The non-perturbative
energy correction, which is left, shows how gluons (and quarks) 
are getting 'dressed' from bare to constituent degrees of freedom.

The solution for the particle number changing part reads    
\begin{eqnarray}
h_{pq}(l)=h_{pq}(0)f(z_{pq}(l))
\,,\label{eq:1.7}\end{eqnarray} 
which shows, that only matrix elements in the band
$|E_p(l)-E_q(l)|\leq 1/\sqrt{l}=\lambda$ survive.
In the third section we consider possible choices for the similarity function.
For example, for the similarity function
\begin{eqnarray}
f(z) = {\rm exp}(-\sqrt{z})
\,,\label{eq:1.8}\end{eqnarray} 
the particle number changing part decays exponentially as $l\rightarrow\infty$
(or $\lambda\rightarrow 0$).
In the case of degenerate eigenvalues of initial Hamiltonian,
$E_p(0)=E_q(0)$, the paritcle number changing part still decays,
but algebraically
\begin{eqnarray}
{\rm exp}(-\sqrt{z})\sim \left( \lambda/\tilde{\mu}\right)^{-
2\tilde{\mu}/\lambda}
\,,\label{eq:1.9}\end{eqnarray} 
due to non-perturbative dependence of energy on flow parameter
$\delta E_{NP}(\lambda)\sim \tilde{\mu} \ln (\lambda^2/\tilde{\mu}^2)$, where
$\tilde{\mu}$ is some energy scale (see the second section).
For the particle number conserving sector one has
\begin{eqnarray}
\frac{dh_{pp'}(l)}{dl}=-\sum_{q}\left(
\frac{dh_{pq}(l)}{dl}\frac{1}{E_p(l)-E_q(l)}h_{qp'}(l) +
h_{pq}(l)\frac{1}{E_{p'}(l)-E_q(l)}\frac{dh_{qp'}(l)}{dl}
\right)
\,,\label{eq:1.10}\end{eqnarray}
and analagously for the $Q$ space. Here $h_{pq}(l)$ is given by Eq. (\ref{eq:1.7}).
When the sectors are assigned as in Eq. (\ref{eq:1.3}), the equation for the diagonal
matrix elements in $P$ space, $p=p'$,
\begin{eqnarray}
\frac{dE_p(l)}{dl} =-\sum_{q}\frac{1}{E_p(l)-E_q(l)}
\frac{d}{dl}\left(
h_{pq}(l)h_{qp}(l)
\right)
\,,\label{eq:1.11}\end{eqnarray} 
provides (after integrating over the flow parameter) 
the gap equation for an effective gluon mass. The equation in $Q$-space
\begin{eqnarray}
\frac{dh_{qq'}(l)}{dl}=-\sum_{p}\left(
\frac{dh_{qp}(l)}{dl}\frac{1}{E_q(l)-E_p(l)}h_{pq'}(l) +
h_{qp}(l)\frac{1}{E_{q'}(l)-E_p(l)}\frac{dh_{pq'}(l)}{dl}
\right)
\,,\label{eq:1.12}\end{eqnarray}
defines an effective $q\bar{q}$ interaction.
The ultimate aim is to solve these equations selfconsistently.
In the next two sections these equations are solved analytically
doing some approximations.

\section{Gluon gap equation}
\label{sec:2}

Integrating flow equations over the flow parameter in one-body sector 
gives gap equations for the effective energies of quark and gluon, 
Eq. (\ref{eq:1.11}). 
Provided the connection between light-front energy and mass
is given
$p^-=\frac{p_{\perp}^2+m^2(l)}{p^+}$
for quark and
$q^-=\frac{q_{\perp}^2+\mu^2(l)}{q^+}$
for gluon,
flow equations for quark and gluon
effective masses are
\begin{eqnarray}
\frac{dm^2(l)}{dl} &=& -(T^aT^a)\int
\frac{dk_1^+d^2k_{1\perp}}{16\pi^3}
g_{q}^2(l)\frac{1}{D_3(l)}
\frac{df^2(D_3(l);l)}{dl}
\frac{\Theta(k_1^+)}{k_1^+}\frac{\Theta(k_2^+)}{k_2^+}
\nonumber\\
&\times&
\bar{u}(p)D_{\mu\nu}(k_1)\gamma^{\mu}
(\not\!k_2+m(l))\gamma^{\nu}u(p)
\delta^{(3)}(p-k_1-k_2)
\,,\label{eq:2.1}\end{eqnarray}
and
\begin{eqnarray}
\frac{d\mu^2(l)}{dl} g_{\mu\nu}\delta_{ab}&=& -{\rm Tr}(T^aT^b)
\int\frac{dk_1^+d^2k_{1\perp}}{16\pi^3}
g_q^2(l)\frac{1}{D_2(l)}
\frac{df^2(D_2(l);l)}{dl}
\frac{\Theta(k_1^+)}{k_1^+}\frac{\Theta(k_2^+)}{k_2^+}
\nonumber\\
&\times&
{\rm Tr}\left( 
\gamma^{\mu}(\not\!k_1+m(l))\gamma^{\nu}(-\not\!k_2+m(l))
\right)
\delta^{(3)}(q-k_1-k_2)
\\
&+&\frac{1}{2}(f^{acd}f^{bdc})
\int\frac{dk_1^+d^2k_{1\perp}}{16\pi^3}
g_g^2(l)\frac{1}{D_1(l)}
\frac{df^2(D_1(l);l)}{dl}
\frac{\Theta(k_1^+)}{k_1^+}\frac{\Theta(k_2^+)}{k_2^+}
\nonumber\\
&\times&
\Gamma^{\mu\sigma\rho}(q,-k_1,-k_2)D_{\sigma\sigma'}(k_1)
\Gamma^{\nu\rho'\sigma'}(-q,k_2,k_1)D_{\rho'\rho}(-k_2)
\delta^{(3)}(q-k_1-k_2)
\nonumber
\,,\label{eq:2.2}\end{eqnarray}
where in the last term the factor $\frac{1}{2}$ is a symmetry factor 
for two-boson states.
Note, that no correction arises to the term $\frac{p_{\perp}^2}{p^+}$
(and $\frac{q_{\perp}^2}{q^+}$) which is protected by the kinematical symmetries;
the total transverse momentum does not appear in a boost invariant expression.
 
The gluon couples to the quark-anti-quark pairs and pairs of gluons
while the quark couples only to the quark-gluon pairs.
Here the energy differences are
\begin{eqnarray}
D_1 &=& q^--k_1^--k_2^-
\nonumber\\
D_2 &=& q^--k_1^--k_2^-
\nonumber\\
D_3 &=& p^--k_1^--k_2^-
\,.\label{eq:2.3}\end{eqnarray}
One should not be confused, that the momenta in different loops are denoted
by the same letters, $k_1$ and $k_2$. 
The trigluon vertex is $(-g_g)\Gamma^{\mu\nu\rho}$, with \cite{BrodskyLepage}
\begin{eqnarray}
\Gamma^{\mu\nu\rho}(p,q,k)=(p-q)^{\rho}g^{\mu\nu}
+(q-k)^{\mu}g^{\rho\nu}+(k-p)^{\nu}g^{\mu\rho}
\,.\label{eq:2.4}\end{eqnarray}
Generally, the quark-gluon coupling, $g_q(l)$, and the trigluon coupling, $g_g(l)$, 
are different from each other (functions of three vertex momenta) for a nonzero flow parameter.
The energy differences depend on the flow parameter $l$
through the masses, i.e.
$D_3(l)=(p_{\perp}^2+m^2(l))/p^+-(k_{1\perp}^2+\mu^2(l))/k_1^+-(k_{2\perp}^2+m^2(l))/k_2^+$.
The polarization sum (in the light-front gauge) is given \cite{BrodskyLepage}
\begin{eqnarray}
D_{\mu\nu}(k) 
= \sum_{\lambda=1,2}\epsilon_{\mu}(\lambda)\epsilon^*_{\nu}(\lambda)
= -g_{\mu\nu} +
\frac{\eta_{\mu}k_{\nu}+\eta_{\nu}k_{\mu}}{k^+}
\,,\label{eq:2.5}\end{eqnarray}
where $k\cdot\epsilon=\eta\cdot\epsilon=0$, and 
the light-front vector $\eta$ is defined as $\eta\cdot k=k^+$.
The sum over helicities for the Dirac spinors is 
\begin{eqnarray}
\sum_{\sigma=\pm 1/2} u(p,\sigma)\bar{u}(p,\sigma) &=&\not\!p+m
\nonumber\\
\sum_{\sigma=\pm 1/2} v(p,\sigma)\bar{v}(p,\sigma) &=&\not\!p-m
\,.\label{eq:2.6}\end{eqnarray} 
We introduce
\begin{eqnarray}
Q^2_1(\lambda) &=& -q^+D_1(\lambda)
\nonumber\\
Q^2_2(\lambda) &=& -q^+D_2(\lambda)
\nonumber\\
Q^2_3(\lambda) &=& -p^+D_3(\lambda)
\,.\label{eq:2.7}\end{eqnarray}
In the light-front frame Eq. (\ref{eq:2.1}) and Eq. (\ref{eq:2.2}) read
\begin{eqnarray}
\frac{dm^2(\lambda)}{d\lambda} &=&
C_f\int_0^1\frac{dx}{x(1-x)}
\int_0^{\infty}\frac{d^2k_{\perp}}{16\pi^3}
g_q^2(\lambda)\frac{1}{Q^2_3(\lambda)}
\frac{df^2(Q^2_3(\lambda);\lambda)}{d\lambda}
\nonumber\\
&\times&
\left(
k_{\perp}^2(\frac{2}{1-x}+\frac{4}{x^2})
+2m^2(\lambda)\frac{x^2}{1-x}
\right)
\,,\label{eq:2.8}\end{eqnarray}
and
\begin{eqnarray}
\frac{d\mu^2(\lambda)}{d\lambda} &=&
2T_fN_f\int_0^1\frac{dx}{x(1-x)}
\int_0^{\infty}\frac{d^2k_{\perp}}{16\pi^3}
g_q^2(\lambda)\frac{1}{Q^2_2(\lambda)}
\frac{df^2(Q^2_2(\lambda);\lambda)}{d\lambda}
\nonumber\\
&\times&
\left(
\frac{k_{\perp}^2+m^2(\lambda)}{x(1-x)}
-2k_{\perp}^2\right)
\nonumber\\
&+& 2C_a\int_0^1\frac{dx}{x(1-x)}
\int_0^{\infty}\frac{d^2k_{\perp}}{16\pi^3}
g_g^2(\lambda)\frac{1}{Q_1^2(\lambda)}
\frac{df^2(Q_1^2(\lambda);\lambda)}{d\lambda}
\nonumber\\
&\times&
\left(
k_{\perp}^2(1+\frac{1}{x^2}+\frac{1}{(1-x)^2})
\right)
\,,\label{eq:2.9}\end{eqnarray}
where
\begin{eqnarray}
Q_1^2(\lambda) &=& \frac{k_{\perp}^2+\mu^2(\lambda)}{x(1-x)}
-\mu^2(\lambda)
\nonumber\\
Q_2^2(\lambda) &=&
\frac{k_{\perp}^2+m^2(\lambda)}{x(1-x)}
-\mu^2(\lambda)
\nonumber\\
Q_3^2(\lambda) &=& \frac{k_{\perp}^2+m^2(\lambda)}{x}
+\frac{k_{\perp}^2+\mu^2(\lambda)}{1-x}
-m^2(\lambda)
\,,\label{eq:2.10}\end{eqnarray}
and we used the connection between the flow parameter, $l$, and the ultraviolet 
cut-off, $\lambda$, as $l=1/\lambda^2$.
Here Casimir operators in fundamental and adjoint representations are, respectively,
$C_f=T^aT^a=(N_c^2-1)/2N_c$ and $C_a\delta_{ab}=f^{acd}f^{bcd}=N_c\delta_{ab}$,
$N_c$ is the number of colors (i.e., $N_c=3$); and
$T_f\delta_{ab}={\rm Tr}(T^aT^b)=\frac{1}{2}\delta_{ab}$.
This system of equations, Eq. (\ref{eq:2.8}) and Eq. (\ref{eq:2.9}), was considered also in \cite{Glazek}.

Generally, it is very difficult to solve both equations 
self-consistently.
One of the reasons is that the equations involve running coupling 
constants, $g_q(\lambda)$ and $g_g(\lambda)$, which depend on $\lambda$
in accordance with renormalization group equations. 
(RG equation for the couplings can be obtained from the flow equations 
for the quark-gluon vertex in the third order and for the trigluon vertex
in the forth order, respectively).  
Also initial conditions for these equations
are not known. In the leading order one can decouple the gap equations 
for quark and gluon effective masses
and the RG equation for the coupling constants
if we neglect all dependences
on the cut-off in the right-hand side of the corresponding flow equations.
In order to take into account these dependences we need to consider 
higher orders.

Below we study the gluon gap equation, that generates an effective gluon mass
which depends on the cut-off $\lambda$. In the next section we show that
the exchange by a gluon with the cut-off dependent mass leads to a confining potential
between static quark and antiquark at large distances.

We integrate the flow equation for the gluon energy, Eq. (\ref{eq:2.9}), neglecting 
cut-off dependence of masses and coupling on the r.h.s. The constant 
of integration is assumed to be a 'physical' mass
(for derivation see \cite{Glazek}).  
We take for an effective gluon mass some value $\tilde{\mu}$.
Gluon gap equation for the abelian part (quark loop) reads
\begin{eqnarray}
\mu^2(\lambda) &=& \tilde{\mu}^2
+2g^2T_fN_f\int_0^1dx\int_0^{\infty}
\frac{d^2k_{\perp}}{16\pi^3}
f(Q_2^2(\lambda)/\lambda^2)
\nonumber\\
&\times&
\left(
\frac{\mu^2(\lambda)(2x^2-2x+1)}{k_{\perp}^2+m^2-x(1-x)\mu^2(\lambda)}
+\frac{2m^2}{k_{\perp}^2+m^2-x(1-x)\mu^2(\lambda)}\right.
\nonumber\\
&+&\left. (-2+\frac{1}{x(1-x)}) \right)
\,,\label{eq:2.11}\end{eqnarray}
where we rescaled the cut-off $\lambda\rightarrow\lambda^2/q^+$.
Similarity function plays the role of UV cut-off in the loop integral.
It also regulates the light-front IR divergences (for $m\neq 0$)
\begin{eqnarray}
&& k^2_{\perp max} = x(1-x)(\lambda^2+\mu^2(\lambda))-m^2
\nonumber\\
&& \frac{m^2}{\lambda^2+\mu^2(\lambda)}\leq x
\leq 1- \frac{m^2}{\lambda^2+\mu^2(\lambda)}
\,.\label{eq:2.12}\end{eqnarray}
One has
\begin{eqnarray}
\mu^2(\lambda) &=& \tilde{\mu}^2
+\frac{g^2T_fN_f}{8\pi^2}
\int_{x_{min}}^{x_{max}}dx \left(
\mu^2(\lambda)(2x^2-2x+1)
\ln\left| \frac{x(1-x)\lambda^2}
{m^2-x(1-x)\mu^2(\lambda)}\right|
\right.
\nonumber\\
&+& \left.
2m^2 \ln\left|\frac{x(1-x)\lambda^2}
{m^2-x(1-x)\mu^2(\lambda)}\right|
+\frac{2}{3}(\lambda^2+\mu^2(\lambda))\right.
\nonumber\\
&-& \left. 2m^2(\ln\frac{\lambda^2+\mu^2(\lambda)}{m^2}-1)
\right)
\,.\label{eq:2.13}\end{eqnarray}

When the renormalization point is taken at
$q^2=0$ and $\tilde{\mu}^2=0$ the gap equation Eq. (\ref{eq:2.11}) 
(and Eq. (\ref{eq:2.13})) 
is reduced to the perturbative case. The perturbation correction is given  
\cite{ZhangHarindranath}
\begin{eqnarray}
\delta\mu^2_{PT}(\lambda)=\frac{g^2T_fN_f}{4\pi^2}\frac{\lambda^2}{3}
\,.\label{eq:2.14}\end{eqnarray}

Non-perturbative solution of the integral equation Eq. (\ref{eq:2.13}) can be 
obtained numerically.
Instead, we solve Eq. (\ref{eq:2.13}) itteratively.
In the leading order 
$\mu^2(\lambda)=\tilde{\mu}^2$.
The next order reads
\begin{eqnarray}
\mu^2(\lambda) &=& \tilde{\mu}^2
+\frac{g^2T_fN_f}{8\pi^2}
\int_{x_{min}}^{x_{max}}dx \left(
\tilde{\mu}^2(2x^2-2x+1)
\ln\left|\frac{x(1-x)\lambda^2}
{m^2-x(1-x)\tilde{\mu}^2}\right|
\right.
\nonumber\\
&+& \left.
2m^2\ln\left|\frac{x(1-x)\lambda^2}
{m^2-x(1-x)\tilde{\mu}^2}\right|
+\frac{2}{3}(\lambda^2+\tilde{\mu}^2)\right.
\nonumber\\
&-& \left. 2m^2(\ln\frac{\lambda^2+\tilde{\mu}^2}
{m^2}-1)
\right)
\,.\label{eq:2.15}\end{eqnarray}
This is equivalent to take the renormalization point 
$q^2=\tilde{\mu}^2$.
Provided that
$m\ll\tilde{\mu}\ll\lambda$, Eq. (\ref{eq:2.15}) is reduced
\begin{eqnarray}
\mu^2(\lambda) &=& \tilde{\mu'}^2
+\frac{g^2T_fN_f}{4\pi^2}\left(
\frac{\lambda^2}{3}
+(\frac{1}{3}+\frac{m^2}{\tilde{\mu}^2}
+2(\frac{m^2}{\tilde{\mu}^2})^2)
\tilde{\mu}^2\ln\frac{\lambda^2}{\tilde{\mu}^2}
\right)
\nonumber\\
&=&\tilde{\mu'}^2
+\delta\mu_{PT}^2(\lambda)
+\delta\mu_{NP}^2(\lambda;\tilde{\mu})
\,.\label{eq:2.16}\end{eqnarray}
We add the finite part (independent on the cut-off $\lambda$) to 
$\tilde{\mu}$,
the result is denoted as $\tilde{\mu}^{\prime}$.
We associate all the terms which depend on $\lambda$ but do not 
depend on $\tilde{\mu}$ with perturbative correction,
denoted as $\delta\mu^2_{PT}(\lambda)$.
The rest, except constant $\tilde{\mu}^{\prime}$,
gives non-perturbative mass correction
$\delta\tilde{\mu}^2_{NP}(\lambda,\tilde{\mu})$.
The perturbative term is given in Eq. (\ref{eq:2.14}).

We consider the non-abelian part (gluon loop), Eq. (\ref{eq:2.9}).
For simplicity we consider exchange with massless gluon; the external 
gluon is put on mass-shell $q^2=\mu^2(\lambda)$, i.e.
the energy denominator is
\begin{eqnarray}
Q_1^{\prime 2}(\lambda) =
\frac{k_{\perp}^2}{x(1-x)}-\mu^2(\lambda)
\,.\label{eq:2.17}\end{eqnarray}
Gap equation (non-abelian part) reads
\begin{eqnarray}
\mu^2(\lambda) &=& \tilde{\mu}^2 
+2g^2C_a\int_0^1dx
\int_0^{\infty}\frac{d^2k_{\perp}}{16\pi^3}
f(Q_1^{\prime 2}(\lambda)/\lambda^2)
\nonumber\\
&\times&
\left(1+\frac{\mu^2(\lambda)x(1-x)}
{k_{\perp}^2-x(1-x)\mu^2(\lambda)}\right)
\left(1+\frac{1}{x^2}+\frac{1}{(1-x)^2}\right)
\,.\label{eq:2.17a}\end{eqnarray}
The range of integration is defined by the similarity function
\begin{eqnarray}
&& k_{\perp max}^2=(\lambda^2+\mu^2(\lambda))x(1-x)
\nonumber\\
&& 0\leq x\leq 1
\,.\label{eq:2.18}\end{eqnarray}
The IR singularities, from $x\rightarrow 0$ and $x\rightarrow 1$,
are not regulated by the UV cut-off for $k_{\perp}$,
since the massless gluon is exchanged. 
We regulate it by principle value prescription
\cite{ZhangHarindranath}
\begin{eqnarray}
\frac{1}{x}\rightarrow\frac{1}{2}\left(\frac{1}{x+i\epsilon_x}
+\frac{1}{x-i\epsilon_x}\right)
\,,\label{eq:2.19}\end{eqnarray}
where $\epsilon_x$ is dimensionless and is boost invariant
(i.e. $\epsilon_x=\epsilon/q^+$).
The result of integration of Eq. (\ref{eq:2.17a}) reads
\begin{eqnarray}
\mu^2(\lambda)=\tilde{\mu}^{2}
+\frac{g^2C_a}{4\pi^2}\left(
(\lambda^2+\mu^2(\lambda))
(\ln\frac{1}{\epsilon_x}-\frac{11}{12})
+\mu^2(\lambda)\ln\frac{\lambda^2}{\mu^2(\lambda)}
(\ln\frac{1}{\epsilon_x}-\frac{11}{12})
\right)
\,.\label{eq:2.20}\end{eqnarray}
The perturbative correction, when $q^2=0$ and $\tilde{\mu}=0$, is given 
\cite{ZhangHarindranath}
\begin{eqnarray}
\delta\mu_{PT}^2=\frac{g^2C_a}{4\pi^2}\lambda^2
(\ln\frac{1}{\epsilon_x}-\frac{11}{12})
\,.\label{eq:2.21}\end{eqnarray}
The itterative solution reads
\begin{eqnarray}
\mu^2(\lambda) &=& \tilde{\mu''}^2
+\frac{g^2C_a}{4\pi^2}\left(
\lambda^2
(\ln\frac{1}{\epsilon_x}-\frac{11}{12})
+\tilde{\mu}^2\ln\frac{\lambda^2}{\tilde{\mu}^2}
(\ln\frac{1}{\epsilon_x}-\frac{11}{12})
\right) 
\nonumber\\
&=&\tilde{\mu''}^2
+\delta\mu^2_{PT}(\lambda)
+\delta\mu^2_{NP}(\lambda;\tilde{\mu}) 
\,,\label{eq:2.22}\end{eqnarray}
the finite constant is denoted as $\tilde{\mu''}$,
and $\tilde{\mu''}\sim\tilde{\mu}$.

We combine quark Eq. (\ref{eq:2.16}) and 
gluon Eq. (\ref{eq:2.22}) loops to give the leading 
order result
\begin{eqnarray}
\mu^2(\lambda) &=& \tilde{\mu}^2
+\frac{g^2}{4\pi^2}
\lambda^2\left( 
C_a(\ln\frac{1}{\epsilon_x}-\frac{11}{12}) 
+T_fN_f\frac{1}{3} \right)
\nonumber\\
&+& \frac{g^2}{4\pi^2}
\tilde{\mu}^2\ln\frac{\lambda^2}{\tilde{\mu}^2}
\left( C_a(\ln\frac{1}{\epsilon_x}-\frac{11}{12})
+T_fN_f(\frac{1}{3}+\frac{m^2}{\tilde{\mu}^2})
\right) 
\nonumber\\
&=&\tilde{\mu}^2
+\delta\mu^2_{PT}(\lambda)
+\delta\mu^2_{NP}(\lambda;\tilde{\mu})
\,,\label{eq:2.23}\end{eqnarray}
where the perturbative term 
\begin{eqnarray}
\delta\mu_{PT}^2(\lambda)=\frac{g^2}{4\pi^2}
\lambda^2\left(C_a(\ln\frac{1}{\epsilon_x}-\frac{11}{12})
+ T_fN_f\frac{1}{3} \right)
\,,\label{eq:2.24}\end{eqnarray}
reproduces the known result for the perturbative mass correction
(without instantaneos graphs) \cite{ZhangHarindranath}.
Here we assumed for the constant term 
$\tilde{\mu'}^2+\tilde{\mu''}^2\sim\tilde{\mu}^2$.

The (non-perturbative) gluon mass correction 
$\delta\mu^2_{NP}(\lambda,\tilde{\mu})$
contains logarithmic IR divergence.
Even by adding instantaneous graphs one can not eliminate severe IR divergences 
that appear in the gluon sector \footnote{
There are also instantaneous diagrams, which arise from the normal-ordering Hamiltonian,
and in principle, they should accompany the generated (dynamical) terms in flow equations.
Following the rules of light-cone perturbation theory \cite{BrodskyLepage},
one can take into account instantaneous graphs by replacing intermediate momenta
in the nominator of dynamical diagrams as
$\tilde{k}=\left( k^+,\sum_{in}k^--\sum_{interm}^{\prime}k^-,k_{\perp} \right)$.
For example, for the quark effective mass, Eq. (\ref{eq:2.1}), in order to add 
instantaneous gluon exchange 
one should make the change
$k_{1\mu}\rightarrow \tilde{k}_{1\mu}=p_{\mu}-k_{2\mu}$;
to add instantaneous quark exchange
$k_{2\mu}\rightarrow \tilde{k}_{2\mu}=p_{\mu}-k_{1\mu}$.
Analagously in gluon sector, Eq. (\ref{eq:2.2}).
}\cite{ZhangHarindranath}.
One may introduce a non-zero mass $\tilde{\mu}$ for gluon in the intermediate state
to regulate these divergences (Appendix A).
But this will cause a mass singularity from massless gluon parameter
$\tilde{\mu}\rightarrow 0$ for the on-shell gluon $q^2=0$
\cite{ZhangHarindranath}.
If gluon mass parameter in intermediate state, introduced as IR regulator,
is the same as the mass of in- and out-going gluon (i.e. $q^2=\tilde{\mu}^2$),
then an effective (non-perturbative) gluon mass vanishes as 
$\tilde{\mu}\rightarrow 0$ (Appendix A).
These problems can be avoided if we introduce a mass scale $u$,
as suggested by Zhang and Harindranath \cite{ZhangHarindranath},
for the minimum cut-off for transverse momentum, $k_{\perp}$. 
This is equivalent to the integration of flow equitions 
for an effective gluon mass
over the flow parameter in finite limits $\{u;\lambda\}$ \cite{Glazek}.
The integration of Eq. (\ref{eq:2.9}) (the non-abelian part) gives 
\begin{eqnarray}
\mu^2(\lambda) &=& \mu^2(u)
+2g^2C_a\int_0^1dx
\int_0^{\infty}\frac{d^2k_{\perp}}{16\pi^3}
(f^2(\tilde{Q}_1^2;\lambda)-f^2(\tilde{Q}_1^2;u))
\nonumber\\
&\times&
\left(1+\frac{\tilde{\mu}^2 x(1-x)}
{k_{\perp}^2-x(1-x)\tilde{\mu}^2+\tilde{\mu}^2}
-\frac{\tilde{\mu}^2}{k_{\perp}^2-x(1-x)\tilde{\mu}^2+\tilde{\mu}^2}\right)
\nonumber\\
&\times&
\left(1+\frac{1}{x^2}+\frac{1}{(1-x)^2}\right)
\,,\label{eq:2.25}\end{eqnarray}
where the renormalization point is $q^2=\tilde{\mu}^2$, and
\begin{eqnarray}
\tilde{Q}_1^2=\frac{k_{\perp}^2+\tilde{\mu}^2}{x(1-x)}-\tilde{\mu}^2
\,.\label{eq:2.26}\end{eqnarray}
The similarity function restricts the transverse momentum to
\begin{eqnarray}
k_{\perp max} &=& (\lambda^2+\tilde{\mu}^2)x(1-x)-\tilde{\mu}^2
\nonumber\\
k_{\perp min} &=& (u^2+\tilde{\mu}^2)x(1-x)-\tilde{\mu}^2
\,,\label{eq:2.27}\end{eqnarray}
and the $x$ range of integration, provided $u\ll\lambda$, is
\begin{eqnarray}
\frac{\tilde{\mu}^2}{u^2+\tilde{\mu}^2}\leq x\leq
1-\frac{\tilde{\mu}^2}{u^2+\tilde{\mu}^2}
\,.\label{eq:2.28}\end{eqnarray}
Integration of Eq. (\ref{eq:2.25}) gives
\begin{eqnarray}
\mu^2(\lambda)-\mu^2(u)=
\frac{g^2C_a}{4\pi^2}\left( \tilde{\mu}^2\ln\frac{\lambda^2}{u^2}
(-\frac{u^2}{\tilde{\mu}^2}
+\ln\frac{u^2}{\tilde{\mu}^2}-\frac{5}{12})
+(\lambda^2-u^2)(\ln\frac{u^2}{\tilde{\mu}^2}-\frac{11}{12})
\right)
\,,\label{eq:2.29}\end{eqnarray}
where $\tilde{\mu}^2\ll u^2$.
The value $\mu^2(u)$ can be found from a renormalization condition
for the 'physical' gluon mass \cite{Glazek}.
The following condition for fitting $\mu^2(u)$ is assumed:
the effective Hamiltonian at the scale $u$ has bosonic eigenstates
with eigenvalues of the form $q^-=\frac{q_{\perp}^2+\tilde{\mu}^2}{q^+}$,
i.e.
\begin{eqnarray}
\frac{q_{\perp}^2+\tilde{\mu}^2}{q^+}\langle q',q\rangle =
\frac{q_{\perp}^2+\mu^2(\lambda)}{q^+}\langle q',q\rangle
-\int^{\lambda} d\lambda' \langle q'|[\eta^{(1)}(\lambda'), H^{(1)}(\lambda')] 
|q\rangle
\,,\label{eq:2.30}\end{eqnarray}
where $\tilde{\mu}$ denotes the 'physical' gluon mass;
$\eta^{(1)}$ is the first order generator which eliminates
the quark gluon (trigluon) coupling constant $H^{(1)}$;
$|q\rangle$ denotes a single effective gluon state with momentum
$q^+$ and $q_{\perp}$, 
$\langle q'|q\rangle =16\pi^3q^+\delta^{(3)}(q'-q)$.
In fact, the initial gap equations, Eq. (\ref{eq:2.11}) and Eq. (\ref{eq:2.17a}), 
were obtained using this renormalization condition, Eq. (\ref{eq:2.30}).
In the order $O(g^2)$ one obtains
\begin{eqnarray}
\tilde{\mu}^2=\mu^2(\lambda)
-\frac{g^2C_a}{4\pi^2}\left( \tilde{\mu}^2\ln\frac{\lambda^2}{\tilde{\mu}^2}
(-\frac{u^2}{\tilde{\mu}^2}
+\ln\frac{u^2}{\tilde{\mu}^2}-\frac{5}{12})
+\lambda^2 (\ln\frac{u^2}{\tilde{\mu}^2}-\frac{11}{12})
\right)
\,.\label{eq:2.31}\end{eqnarray}
So,
\begin{eqnarray}
\mu^2(u)=\tilde{\mu}^2
+\frac{g^2C_a}{4\pi^2}\left( \tilde{\mu}^2\ln\frac{u^2}{\tilde{\mu}^2}
(-\frac{u^2}{\tilde{\mu}^2}
+\ln\frac{u^2}{\tilde{\mu}^2}-\frac{5}{12})
+u^2 (\ln\frac{u^2}{\tilde{\mu}^2}-\frac{11}{12})
\right)
\,.\label{eq:2.32}\end{eqnarray}
Therefore, the effective gluon mass in the interacting Hamiltonian 
at the scale $\lambda$ is given
\begin{eqnarray}
\mu^2(\lambda)=\tilde{\mu}^2
+\frac{g^2C_a}{4\pi^2}\left( \tilde{\mu}^2\ln\frac{\lambda^2}{\tilde{\mu}^2}
(-\frac{u^2}{\tilde{\mu}^2}
+\ln\frac{u^2}{\tilde{\mu}^2}-\frac{5}{12})
+\lambda^2 (\ln\frac{u^2}{\tilde{\mu}^2}-\frac{11}{12})
\right)
\,,\label{eq:2.33}\end{eqnarray}
where parameter $u^2$ plays the role of IR regulator. 
This equation can be compared to Eq. (\ref{eq:2.22}).

To be consistent, we integrate also the quark loop 
over the flow parameter in finite limits, though in this case
the IR singularities are regulated by the nonzero quark mass, $m\neq 0$.
In full analogy with nonabelian case, one has from Eq. (\ref{eq:2.9}) 
(the abelian part)
\begin{eqnarray}
\mu^2(\lambda)-\mu^2(u)=
\frac{g^2T_fN_f}{4\pi^2}\left(
\frac{1}{3}
\tilde{\mu}^2\ln\frac{\lambda^2}{u^2}
+m^2\ln\frac{\lambda^2}{u^2}
+\frac{1}{3}(\lambda^2-u^2)
\right)
\,.\label{eq:2.34}\end{eqnarray}
Therefore
\begin{eqnarray}
\mu^2(\lambda)=\tilde{\mu}^2
+\frac{g^2T_fN_f}{4\pi^2}\left(
\frac{1}{3}
\tilde{\mu}^2\ln\frac{\lambda^2}{\tilde{\mu}^2}
+m^2\ln\frac{\lambda^2}{\tilde{\mu}^2}
+\frac{1}{3}\lambda^2
\right)
\,,\label{eq:2.35}\end{eqnarray}
that one can compare to Eq. (\ref{eq:2.16}).
We combine the abelian, Eq. (\ref{eq:2.35}), and nonabelian, Eq. (\ref{eq:2.33}), terms 
\begin{eqnarray}
\mu^2(\lambda) &=& \tilde{\mu}^2
+\frac{g^2}{4\pi^2} \lambda^2 
 \left(C_a (\ln\frac{u^2}{\tilde{\mu}^2}-\frac{11}{12})
+T_fN_f\frac{1}{3}\right)
\nonumber\\
&+&
\frac{g^2}{4\pi^2} \tilde{\mu}^2\ln\frac{\lambda^2}{\tilde{\mu}^2}
\left( C_a (-\frac{u^2}{\tilde{\mu}^2}
+\ln\frac{u^2}{\tilde{\mu}^2}-\frac{5}{12})
+T_fN_f(\frac{1}{3}+\frac{m^2}{\tilde{\mu}^2})\right)
\nonumber\\
&=& \tilde{\mu}^2+\delta\mu_{PT}^2(\lambda)
+\delta\mu_{NP}^2(\lambda,\tilde{\mu},u)
\,.\label{eq:2.36}\end{eqnarray}
The sum of constant terms in Eq. (\ref{eq:2.35}) and Eq. (\ref{eq:2.33}) is of order 
of the mass parameter $\tilde{\mu}^2$.
Here the perturbative term is
\begin{eqnarray}
\delta\mu_{PT}^2(\lambda)=\frac{g^2}{4\pi^2}
\lambda^2 \left(
C_a (\ln\frac{u^2}{\tilde{\mu}^2}-\frac{11}{12})
+T_fN_f\frac{1}{3}\right)
\,.\label{eq:2.37}\end{eqnarray}
Note, that for $u\gg\tilde{\mu}$ the nonperturbative mass correction
in Eq. (\ref{eq:2.36}) changes the sign as compared with nonregulated mass correction
Eq. (\ref{eq:2.23}), since the intermediate gluon with nonzero mass contributes
negative term (see Appendix A or Eq. (\ref{eq:2.25})). In fact,
QCD with nonzero gluon mass resembles Yukawa theory.
It was shown for Yukawa theory, that the leading correction
proportional to $\lambda^2$ and the logarithmic correction
$\sim \ln\lambda^2/\tilde{\mu}^2$ appear with opposite signs \cite{Glazek}.

We introduce the mass counterterm to renormalize the Hamiltonian perturbatively
in the second order
\begin{eqnarray}
m_{CT}^2=-\delta\mu_{PT}^2(\lambda=\Lambda\rightarrow\infty)
\,,\label{eq:2.38}\end{eqnarray}
that removes the leading cut-off dependence.
The rest is the (nonperturbative) effective gluon mass
\begin{eqnarray}
\mu_{NP}^2(\lambda)=\tilde{\mu}^2
-\sigma(\tilde{\mu},u)\ln\frac{\lambda^2}{\tilde{\mu}^2}
\,,\label{eq:2.39}\end{eqnarray}
where we introduced 
\begin{eqnarray}
\sigma(\tilde{\mu},u)= -\frac{g^2}{4\pi^2}
\tilde{\mu}^2 \left( C_a (-\frac{u^2}{\tilde{\mu}^2}
+\ln\frac{u^2}{\tilde{\mu}^2}-\frac{5}{12})
+T_fN_f(\frac{1}{3}+\frac{m^2}{\tilde{\mu}^2}) \right)
\,.\label{eq:2.40}\end{eqnarray}
The limit of the $\sigma(\tilde{\mu},u)$ as
$\tilde{\mu}\rightarrow 0$ is finite, and is equal  
\begin{eqnarray}
\sigma=\lim_{\tilde{\mu}\to 0}\sigma(\tilde{\mu},u)
=\frac{g^2C_a}{2\pi^2}u^2
\,,\label{eq:2.41}\end{eqnarray}
that plays the role of the string tension between quark and antiquark
(see next section).
One should not be confused that the string tension is proportional
to the coupling constant. In Appendix A we show, that by the proper regularization
of IR divergences the string tension has a pure non-perturbative form.
In Eq. (\ref{eq:2.40}) we can remove the coupling constant by rescaling
$u^2\rightarrow u^2/g^2$, where $g$ is the renormalized coupling constant.

In Eq. (\ref{eq:2.39}) for $\lambda\ll\tilde{\mu}$ the effective (nonperturbative) gluon mass
equals the 'physical' gluon mass, mass parameter $\tilde{\mu}$.
For $\lambda\gg\tilde{\mu}$ there is the nonperturbative correction 
to the mass $\tilde{\mu}$, given by the second term, which describes
'dressing' of gluon.

\section{Confinement}
\label{sec:3}

Eliminating the quark gluon coupling one obtains
an effective interaction between quark and antiquark, Eq. (\ref{eq:1.12}). 
The effective interaction between electron and positron
in the light-front gauge was obtained
in the previous work\footnote{
The difference in the prefactor between \cite{GubankovaWegner}
and Eq. (\ref{eq:3.1}) is $\times \left(\frac{1}{16\pi^3}\right)$, which
comes from the light-front normalization in the bound state
integral equation $\sim\int \frac{d^2k_{\perp}}{16\pi^3}$. 
}
\cite{GubankovaWegner}
\begin{eqnarray}
V_{e\bar{e}}=-4\pi^2\alpha_{em}
\langle\gamma^{\mu}\gamma^{\nu}\rangle
B_{\mu\nu}
\,,\label{eq:3.1}\end{eqnarray}
where the current-current term
in exchange channel is given
\begin{eqnarray}
\langle\gamma^{\mu}\gamma^{\nu}\rangle
=\frac{\left( \bar{u}(p'_1,\lambda'_1)\gamma^{\mu}u(p_1,\lambda_1)\right)
\left(\bar{v}(p_2,\lambda_2)\gamma^{\nu}v(p'_2,\lambda'_2)\right)}
{\sqrt{xx'(1-x)(1-x')}}
\,,\label{eq:3.2}\end{eqnarray}
where $x=p_1^+/P^+$ is the longitudinal momentum fraction.
The energy transfers along electron and positron lines are given,
respectively
\begin{eqnarray}
D_1 &=& p_1^--p_1^{\prime -}-q^-
\nonumber\\
D_2 &=& p_2^{\prime -}-p_2^--q^-
\,,\label{eq:3.3}\end{eqnarray}
where $q=p_1-p'_1=(q^+,q_{\perp})$ is the photon momentum.
The energy differences Eq. (\ref{eq:3.3}) are related to the four-momentum transfers
\begin{eqnarray}
Q_1^2 &=& -(p_1-p'_1)^2= -q^+D_1\nonumber\\
Q_2^2 &=& -(p'_2-p_2)^2= -q^+D_2
\,.\label{eq:3.4}\end{eqnarray}
The following combinations are useful
\begin{eqnarray}
Q^2 &=& (Q_1^2+Q_2^2)/2
\nonumber\\
\delta Q^2 &=& (Q_1^2-Q_2^2)/2
\,.\label{eq:3.4a}\end{eqnarray}
We introduce also the energy differences in $t$-channel
\begin{eqnarray}
D'_1 &=& p_1^-+p_2^--(p_1+p_2)^-
\nonumber\\
D'_2 &=& p_1^{\prime -}+p_2^{\prime -}-(p'_1+p'_2)^-
\,,\label{eq:3.5}\end{eqnarray}
and will use them later. They are related to the invariant mass-squares
of the initial and final states
\begin{eqnarray}
M_1^2 &=& (p_1+p_2)^2= p^+D'_1\nonumber\\
M_2^2 &=& (p'_1+p'_2)^2= p^+D'_2
\,,\label{eq:3.6}\end{eqnarray}
also
\begin{eqnarray}
M^2 &=& (M_1^2+M_2^2)/2 
\nonumber\\
\delta M^2 &=& (M_1^2-M_2^2)/2
\,.\label{eq:3.6a}\end{eqnarray}
The tensor part $B_{\mu\nu}$ of the effective interaction Eq. (\ref{eq:3.1})
includes two terms
\begin{eqnarray}
B_{\mu\nu}=B_{\mu\nu}^{gen}
+B_{\mu\nu}^{inst}
\,.\label{eq:3.7}\end{eqnarray}
The first one is generated
by flow equations in the second order of perturbation theory
\begin{eqnarray}
B_{\mu\nu}^{gen}=D_{\mu\nu}(q)\left(
\frac{\Theta_1}{Q_1^2}+\frac{\Theta_2}{Q_2^2}
\right)
\,,\label{eq:3.8}\end{eqnarray}
and describes the dynamical photon exchange between electron
and positron in s-channel.
Here the polarization sum $D_{\mu\nu}$ is given in Eq. (\ref{eq:2.5}); 
and the energy denominators are given in Eq. (\ref{eq:3.4}).
The $\Theta$-factor is defined
\begin{eqnarray}
\Theta_1=\int_0^{\infty}d\lambda^2
\frac{df(Q_1^2/\lambda^2)}{d\lambda^2}
f(Q_2^2/\lambda^2)
\,,\label{eq:3.9}\end{eqnarray}
where in order to preserve the boost invariance
the cut-off is given in units of longitudinal exchange momentum, i.e.
$\lambda^2/q^+$.
The meaning of the $\Theta$-factor can be interpreted, if we consider 
the same integration as in Eq. (\ref{eq:3.9}) in finite limits
$\Theta(\lambda_0)=\int_{\lambda_0}^{\infty}d\lambda^2(df_1/d\lambda^2)f_2$.
The generated interaction, Eq. (\ref{eq:3.8}) with $\Theta(\lambda_0)$-factors,
appears when high-energy modes are eliminated, since 
$\Theta(\lambda_0)$-factor allows only momenta $Q_1^2\geq\lambda_0^2$
(and $Q_2^2\geq\lambda_0^2$).

The second term in Eq. (\ref{eq:3.7}) is the instantaneous interaction,
which comes from fixing of the light-front gauge \cite{BrodskyLepage}
\begin{eqnarray}
B_{\mu\nu}^{inst}=\frac{\eta_{\mu}\eta_{\nu}}{q^{+2}}
\,.\label{eq:3.10}\end{eqnarray}
The sum of the dynamical and instantaneous terms is given by \cite{GubankovaWegner}
\begin{eqnarray}
B_{\mu\nu}=g_{\mu\nu}
\left(\frac{\Theta_1}{Q_1^2}+\frac{\Theta_2}{Q_2^2}
\right)
+\eta_{\mu}\eta_{\nu}
\frac{\delta Q^2}{q^{+2}}
\left(\frac{\Theta_1}{Q_1^2}-\frac{\Theta_2}{Q_2^2}
\right)
\,,\label{eq:3.11}\end{eqnarray}
where $\delta Q^2$ is given in Eq. (\ref{eq:3.4a}). 
In the light-front frame the energy transfers, Eq. (\ref{eq:3.4}), read
\begin{eqnarray}
Q_1^2 &=& \frac{(x'k_{\perp}-xk'_{\perp})^2+m^2(x-x')^2}{xx'}
\nonumber\\
Q_2^2 &=& Q_1^2 |_{x\rightarrow x';(1-x)\rightarrow (1-x') }
\,,\label{eq:3.12}\end{eqnarray}  
that are always positive $(\geq 0 )$.

We generalize the expression for an effective $e\bar{e}$-interaction, 
Eq. (\ref{eq:3.1}),
to the case of QCD with a nonzero gluon mass.
We simulate an effective interaction in QCD between quark and antiquark
as an exchange of non-perturbative gluon (gluon flux) with
a nonzero effective mass (effective energy),
which evolves with the cut-off $\lambda$ according to RG equation.
The four momentum transfers read
\begin{eqnarray}
Q_1^2(\lambda) &=& Q_1^2+\mu^2_{NP}(\lambda)
\nonumber\\
Q_2^2(\lambda) &=&
Q_1^2(\lambda)|_{x\rightarrow x';(1-x)\rightarrow (1-x') }
\,,\label{eq:3.13}\end{eqnarray}  
where $Q_i^2$ are given in Eq. (\ref{eq:3.12}); and the nonperturbative
effective mass, as obtained in the previous section Eq. (\ref{eq:2.39}), is
\begin{eqnarray}
\mu^2_{NP}(\lambda)=\tilde{\mu}^2-\sigma(\tilde{\mu},u)
\ln\frac{\lambda^2}{\tilde{\mu}^2}
\,.\label{eq:3.14}\end{eqnarray}
To reflect the phenomenological dependence of the effective gluon mass
on the momentum we use the following parametrization
\begin{eqnarray}
&& Q_i^2(\lambda) = \tilde{Q}_i^2
-\tilde{\sigma}\ln\frac{\lambda^2}{\tilde{Q}_i^2}
\nonumber\\
&& \tilde{Q}_i^2 = Q_i^2+\tilde{\mu}^2
\nonumber\\
&& \tilde{\sigma} = \sigma(\tilde{\mu},u)
\,,\label{eq:3.15}\end{eqnarray}
which holds for 
$Q_i^2\leq \tilde{\mu}^2$ and $\tilde{Q}_i^2\leq\lambda^2$.
In fact it does not change the result
for an effective $q\bar{q}$-interaction
what kind of parametrization to use,Eq. (\ref{eq:3.13}) with Eq. (\ref{eq:3.14}) 
or Eq. (\ref{eq:3.15}).

In QCD we take into account the dependence of four-monetum transfers along 
quark and antiqurk lines on the cut-off.
In full analogy with QED, the effective quark-antiquark interaction
reads
\begin{eqnarray}
V_{q\bar{q}}=-Const\langle\gamma^{\mu}\gamma^{\nu}\rangle
\tilde{B}_{\mu\nu}
\,,\label{eq:3.16}\end{eqnarray}
where instead of coupling $\alpha_{em}$ some constant term $Const$
is introduced.
At the end of calculations we fit $Const$ and $\sigma$ 
to reproduce the correct coefficients
for the short-range and long-range parts of $q\bar{q}$-potential.
Here $\tilde{B}_{\mu\nu}$ includes the dynamical and instantaneous 
gluon exchange. Eliminating by flow equations the quark-gluon coupling,
where gluon has an effective cut-off dependent mass, one obtains
the dynamical (generated) interaction
\begin{eqnarray}
\tilde{B}_{\mu\nu}^{gen}=D_{\mu\nu}(q)\left(I_1+I_2\right)
\,,\label{eq:3.17}\end{eqnarray} 
where the integral term is given by
\begin{eqnarray}
I_1=\int_0^{\infty}d\lambda^2\frac{1}{Q_1^2(\lambda)}
\frac{df(Q_1^2(\lambda)/\lambda^2)}{d\lambda^2}
f(Q_2^2(\lambda)/\lambda^2)
\,,\label{eq:3.18}\end{eqnarray}
with $Q_i^2(\lambda)$ defined in Eq. (\ref{eq:3.15}).

In order to obtain the instantaneous interaction one should modify
the QCD Hamiltonian in the light-front gauge for 
the case of nonzero gluon mass. Instead, we use the same rules
to calculate an effective $q\bar{q}$-interaction as in 
the perturbative case for QED \cite{GubankovaWegner}.
Applying flow equations to the light-front QED Hamiltonian,
one can show, that the instantaneous propagator, 
fermion $\gamma^+/2q^+$ and gluon $\eta^{\mu}\eta^{\nu}/q^{+ 2}$, 
can be absorbed into the regular propagator,
$(\not\!q+m)$ and $D_{\mu\nu}(q)$ respectively, 
by replacing $q$,
the momentum associated with the line, by
\begin{eqnarray}
\tilde{q}^{(i)} =
\left( q^+,\sum_{in}q^--\sum_{interm}^{\prime}q^-
\pm\frac{1}{2}
(\sum_{out}q^--\sum_{in}q^-), q_{\perp} \right)
\,,\label{eq:3.19}\end{eqnarray}
in the numerator for those diagrams in which the fermion or gluon 
propagates are only over a single time interval. 
Here $\sum_{in}$ ($\sum_{out}$) denotes summation over all inital 
(final) particles in the diagram, while $\sum_{interm}^{\prime}$ 
denotes summation over all particles in the intermediate state other 
than the particle of interest.
When energy is conserved, 
$\sum_{in}q^- = \sum_{out}q^-$, we recover
the rules of light-front perturbation theory as formulated
by Brodsky and Lepage \cite{BrodskyLepage}.
In Eq. (\ref{eq:3.19}) for index $i=1$ the sign is plus, for $i=2$ is minus.
In order to absorb the instantaneous term Eq. (\ref{eq:3.10})
and obtain an effective interaction Eq. (\ref{eq:3.11}),
one should do the replacement in polarization sum 
of the dynamical term Eq. (\ref{eq:3.8})
\begin{eqnarray}
\sum_{i=1,2}D_{\mu\nu}(q)\frac{\Theta_i}{Q_i^2}
\rightarrow
\sum_{i=1,2}D_{\mu\nu}(\tilde{q}^{(i)})
\frac{\Theta_i}{Q_i^2}
\,,\label{eq:3.20}\end{eqnarray}
where $\tilde{q}^{(i)}$ are given in Eq. (\ref{eq:3.19}).

Following this rule also for QCD, we obtain from 
the dynamical interaction Eq. (\ref{eq:3.17}) an effective $q\bar{q}$-interaction
Eq. (\ref{eq:3.16}), with
\begin{eqnarray}
\tilde{B}_{\mu\nu}=g_{\mu\nu}
\left(I_1+I_2\right)
+\eta_{\mu}\eta_{\nu}
\frac{\delta Q^2}{q^{+2}}
\left(I_1-I_2\right)
\,,\label{eq:3.21}\end{eqnarray}
where the integral terms $I_i$ are defined in Eq. (\ref{eq:3.18}).

Similarity function is a function of the break 
$Q^2(\lambda)/\lambda^2$, where
four momentum transfer $Q^2(\lambda)$
introduces implicit dependence on the cut-off, Eq. (\ref{eq:3.15}). In this case the 
integral term, Eq. (\ref{eq:3.18}), is reduced  
\begin{eqnarray}
I_1=\int_0^{\infty}d\left(\frac{1}{\lambda^2}\right)
\frac{df(z_1)}{dz_1}f(z_2)
\left(1+\frac{\tilde{\sigma}}{Q_1^2(\lambda)}
\right)
\,,\label{eq:3.22}\end{eqnarray}
where $z_1=Q_1^2(\lambda)/\lambda^2$ and 
$z_2=Q_2^2(\lambda)/\lambda^2$.
The first term, unit, describes perturbative one-gluon exchange
(analog of photon exchange in QED);
the second term arises from the dependence of effective gluon mass
on the cut-off and defines the long-range part of the effective interaction,
since it is more singular than the first term.
Provided the properties for similarity function as in Eq. (\ref{eq:1.6}), 
the integral factor $I_i$
saturates by the values
$z_1\sim 1$ and $z_2\sim 1$,
i.e. an effective range of integration is
$0\leq\lambda^2\leq (\lambda_0^2$ and $\lambda_0^{\prime 2})$
where $\lambda_0^2\sim \tilde{Q}_1^2$ and $\lambda_0^{\prime 2}\sim\tilde{Q}_2^2$.
In this range of $\lambda$
the four-momentum transfers do not depend on the cut-off, i.e.
from Eq. (\ref{eq:3.15}) 
$Q_1^2(\lambda_0)\sim\tilde{Q}_1^2$
and  $Q_2^2(\lambda'_0)\sim\tilde{Q}_2^2$.
This enables to estimate the integral factor $I_i$ analytically.
One can approximate the integral factor provided the following relation holds
$\tilde{Q}_1^2\sim \tilde{Q}_2^2$, which is considered below.
 
We consider different similarity functions: 
exponential, Gaussian and sharp cut-offs.
The corresponding integral factors read
\begin{eqnarray} 
&& f =
{\rm exp}\left( -Q^2(\lambda)/\lambda^2\right),
\hspace{1cm}
I_1=\frac{1}{\tilde{Q}_1^2+\tilde{Q}_2^2}
\left(1+\frac{\tilde{\sigma}}{\tilde{Q}_1^2}
\right) 
\nonumber\\
&& f =
{\rm exp}\left(-Q^4(\lambda)/\lambda^4\right),
\hspace{1cm}
I_1=\frac{\tilde{Q}_1^2}{\tilde{Q}_1^4+\tilde{Q}_2^4}
\left(1+\frac{\tilde{\sigma}}{\tilde{Q}_1^2}
\right) 
\nonumber\\
&& f =
\theta(\lambda^2-Q^2(\lambda)),
\hspace{1cm}
I_1=\frac{\theta(\tilde{Q}_1^2-\tilde{Q}_2^2)}
{\tilde{Q}_1^2}
\left(1+\frac{\tilde{\sigma}}{\tilde{Q}_1^2}
\right) 
\,,\label{eq:3.23}\end{eqnarray}
where $\tilde{Q}_i^2$ are given in Eq. (\ref{eq:3.15}).
We define
\begin{eqnarray}
\tilde{Q}^2 &=& (\tilde{Q}_1^2+\tilde{Q}_2^2)/2
= Q^2+\tilde{\mu}^2
\nonumber\\
\delta \tilde{Q}^2 &=& (\tilde{Q}_1^2-\tilde{Q}_2^2)/2
=\delta Q^2
\,,\label{eq:3.24}\end{eqnarray}
with $Q^2$ and $\delta Q^2$ given by Eq. (\ref{eq:3.4a}).   
The effective interaction between
quark and antiquark, Eq. (\ref{eq:3.21}), for the three choices of similarity 
function reads 
\begin{eqnarray}
B_{\mu\nu} &=& g_{\mu\nu}\left(
\frac{1}{\tilde{Q}^2}
+\frac{\tilde{\sigma}}{\tilde{Q}^4}\right)
+\left(\frac{g_{\mu\nu}}{\tilde{Q}^2}
-\frac{\eta_{\mu}\eta_{\nu}}{q^{+2}}\right)
\frac{\tilde{\sigma}}{\tilde{Q}^2}
\frac{\delta\tilde{Q}^4}{\tilde{Q}^4-\delta\tilde{Q}^4} 
\nonumber\\
B_{\mu\nu} &=& g_{\mu\nu}\left(
\frac{1}{\tilde{Q}^2}
+\frac{\tilde{\sigma}}{\tilde{Q}^4}\right)
-\left(\frac{g_{\mu\nu}}{\tilde{Q}^2}
(1+\frac{\tilde{\sigma}}{\tilde{Q}^2})
-\frac{\eta_{\mu}\eta_{\nu}}{q^{+2}}\right)
\frac{\delta\tilde{Q}^4}{\tilde{Q}^4+\delta\tilde{Q}^4} 
\nonumber\\
B_{\mu\nu} &=& g_{\mu\nu}\left(
\frac{1}{\tilde{Q}^2}
+\frac{\tilde{\sigma}}{\tilde{Q}^4}\right)
-\left(
\frac{g_{\mu\nu}}{\tilde{Q}^2}
(1+\frac{\tilde{\sigma}}{\tilde{Q}^2}
(1+\frac{\tilde{Q}^2}
{\tilde{Q}^2+\left|\delta\tilde{Q}^2\right|}))\right.
\nonumber\\
&-& \left.\frac{\eta_{\mu}\eta_{\nu}}{q^{+2}}
(1+\frac{\tilde{\sigma}}{\tilde{Q}^2}
\frac{\tilde{Q}^2}
{\tilde{Q}^2+\left|\delta\tilde{Q}^2\right|})
\right)
\frac{\left|\delta\tilde{Q}^2\right|}
{\tilde{Q}^2+\left|\delta\tilde{Q}^2\right|} 
\,,\label{eq:3.25}\end{eqnarray}
where we defined $\tilde{Q}^4=(\tilde{Q}^2)^2$
and $\delta\tilde{Q}^4=(\delta\tilde{Q}^2)^2$.
For $\delta\tilde{Q}^2\ll\tilde{Q}^2$ the leading effective interaction
in all three cases is the same and is given by the first term of Eq. (\ref{eq:3.25})
\begin{eqnarray}
V_{q\bar{q}}=-Const\langle\gamma^{\mu}\gamma_{\mu}\rangle
\left(\frac{1}{\tilde{Q}^2}+\frac{\tilde{\sigma}}{\tilde{Q}^4}
\right) +
O\left( (\frac{\delta Q^2}{Q^2}) \right)
\,.\label{eq:3.26}\end{eqnarray}
Gluon with an effective mass parameter $\tilde{\mu}$
mediates interaction between quark and antiquark at the distances 
$r\sim 1/\tilde{\mu}$.

We define the resulting $q\bar{q}$-effective interaction in the limit
of the gluon mass parameter $\tilde{\mu}\rightarrow 0$.
In this limit the average four-momentum transfer $\tilde{Q}^2$,
Eq. (\ref{eq:3.24}), and the string tension 
$\tilde{\sigma}=\sigma(\tilde{\mu},u)$,
Eq. (\ref{eq:2.40}), are given 
\begin{eqnarray}
&& \lim_{\tilde{\mu}\rightarrow 0}\tilde{Q}^2 = Q^2
\nonumber\\ 
&& \lim_{\tilde{\mu}\rightarrow 0}\tilde{\sigma} = \sigma
\,.\label{eq:3.27}\end{eqnarray}  
From Eq. (\ref{eq:3.26}) the leading behavior reads
\begin{eqnarray}
V_{q\bar{q}}=-\langle\gamma^{\mu}\gamma_{\mu}\rangle
\left(C_f\alpha_s\frac{4\pi}{Q^2}+\sigma\frac{8\pi}{Q^4}
\right) 
\,,\label{eq:3.28}\end{eqnarray}
which includes Coulomb and confining interactions. We show this 
explicitly below. Here $C_f=T^aT^a=(N_c^2-1)/2N_c$.
Here we restored the correct prefactors before the both terms,
using the freedom to fit the overall constant, $Const$,
and $\sigma$ term which is proportional to the scale $u^2$, 
$\sigma\sim u^2$ in Eq. (\ref{eq:2.41}).

We express the effective $q\bar{q}$-interaction Eq. (\ref{eq:3.28}) in the instant frame.
Instead of the light-front frame we use the instant parametrization
$p(p^+,k_{\perp})\rightarrow (p_z,k_{\perp})$, where the connection between 
the light-front $x$ and z-component of momentum is given
\begin{eqnarray}
x=\frac{1}{2}\left(1+\frac{p_z}{\sqrt{\vec{p}^{\,2}+m^2}} \right)
\,.\label{eq:3.29}\end{eqnarray}
In the instant frame the four momenta read
\begin{eqnarray}
Q^2 &=& \vec{q}^{\,2}-p_zp_z^{\prime}
\frac{(M_1-M_2)^2}{M_1M_2}
\nonumber\\
\delta Q^2 &=& \left( \frac{p_z}{M_1}-\frac{p_z^{\prime}}{M_2}
\right)\delta M^2
\,,\label{eq:3.30}\end{eqnarray}
where $\vec{q}=\vec{p}-\vec{p'}=(q_z,q_{\perp})$
is the three momentum transfer of the gluon. Here
$M_1$ and $M_2$ are the total energies of inital and final states,
defined in Eq. (\ref{eq:3.6}) and the energy difference $\delta M^2$, 
defined in Eq. (\ref{eq:3.6a}), shows the 'off-shellness' of the process.
In the instant frame one has
\begin{eqnarray}
M_1^2 &=& 4(\vec{p}^{\,2}+m^2)
\nonumber\\
M_2^2 &=& 4(\vec{p'}^2+m^2)
\,.\label{eq:3.31}\end{eqnarray}
that enters Eq. (\ref{eq:3.30}).
As $\delta Q^2\rightarrow 0$ the effective interaction
has singularity at $Q^2\rightarrow \vec{q}^{\,2}\rightarrow 0$.
The Fourier transform (with respect to $\vec{q})$ reads 
\begin{eqnarray}
V_{q\bar{q}}=\langle\gamma^{\mu}\gamma_{\mu}\rangle
\left(-C_f\frac{\alpha_s}{r}+\sigma\cdot r
\right)
\,,\label{eq:3.32}\end{eqnarray}
which is the sum of the Coulomb and confining interactions.
Though we were working in the light-front formalism
the result for the leading effective interaction is rotational invariant.

There are corrections $O\left(\frac{\delta Q^2}{Q^2}\right)$ 
to the leading effective interaction, Eq. (\ref{eq:3.28}) 
(or Eq. (\ref{eq:3.32})).
These corrections depend on the direction from which $\vec{q}$
approaches zero. For sufficiently smooth similarity functions $f(z)$,
as exponential and Gaussian cut-offs, the effective interaction
does not contain collinear singularity ($\sim 1/q^+$).
Thus the interaction becomes only singular if $\vec{q}$ 
approaches zero where it diverges like $1/\vec{q}^{\,2}$ (Coulomb singularity)
and $1/\vec{q}^{\,4}$ (confining singularity). However this is not true
for the sharp cut-off, where $\eta_{\mu}\eta_{\nu}$ term
diverges like $1/q^+$. For a smooth cut-off 
the collinear singularity disappears (cancels completely)
and only the rotational invariant 
part of the effective interaction survives 
in the limit $\vec{q}\rightarrow 0$.

\section{Conclusions}

We suggested a possible scenery of confinement in the light-front QCD,
basing on the method of flow equations. Flow equations operate 
in terms of 'physical' (dynamical) degrees of freedom, which are getting
'dressed' through the non-perturbative renormalization of the canonical
QCD Hamiltonian.

Integrating flow equation over the flow parameter in one gluon sector
we obtained the gluon gap equation. It was solved, given an arbitrary
mass parameter $\tilde{\mu}$ in the renormalization point,
for the perturbative and non-perturbative gluon mass corrections.
Performing perturbative renormalization, the perturbative correction
is absorbed by the second order mass counterterm.
Severe collinear IR divergences, which arise in the gluon sector
from the non-abelian gluon interactions, were regulated by introducing
an additional cut-off $u$ for the transverse momentum $k_{\perp}$ in IR region.
This is equivalent to integration of gluon flow equation in the finite limits,
from the bare cut-off $\lambda=\Lambda_{UV}$ down to the hadron scale $u$,
with $u\ll\Lambda_{UV}$. The result is the nonperturbative effective
gluon mass. 

Eliminating the quark gluon coupling by flow equations, one obtains
an effective interaction between quark and antiquark.
In this approach the exchange with the dynamical gluon mode
between the probe quarks, where the effective gluon mass evolves
with RG equations, gives rise to the effective $q\bar{q}$-interaction
which exibits Coulomb and confining singularities.
The cut-off $u$, which regulates IR divergence, sets up a scale
for the long-range part of interaction: it defines the string tension
of confining interaction, $\sigma\sim u^2$. This suggests
some relation between the zero modes of $A^+$ and confinement mechanism
in the light-front formalism.

For a smooth cut-off function the collinear singularity is canceled,
and the leading effective interaction has rotationally invariant form.
It is not true for the sharp cut-off.

The ultimate goal of the study is to solve the chain of flow equations
in different sectors selfconsistently.
As was shown in this work, even an approximate 
solution of the gluon gap equation together with the flow equation 
for an effective interaction between probe quarks
may provide an understanding of confinement. The next step
is to include dynamical quark degrees of freedom, and to address
in this formalism the problem of chiral symmetry breaking in QCD. 

This study shows, that in the light-front quantization it is possible
to isolate the degrees of freedom that are responsible 
for the long-range properties of QCD, and obtain some insight
into the non-perturbative QCD phenomena. Probably the light-front formalism
is the most suitable frame to try solving selfconsistently the system 
of flow equations on computer.

\newpage
\appendix
\section{IR regularization via an effective gluon mass}
\label{sec:4}

We consider correction to the effective gluon mass which arise from 
the non-abelian part. The IR singular behavior is regulated by the 
same mass parameter. From Eq. (\ref{eq:2.9}) (the non-abelian part)
one has 
\begin{eqnarray}
\mu^2(\lambda) &=& \tilde{\mu}^2
+2g^2C_a\int_0^{1}dx \int_0^{\infty}
\frac{d^2k_{\perp}}{16\pi^3}
f(Q_1^2(\lambda)/\lambda^2)
\nonumber\\
&\times&
\left(
1+\frac{\mu^2(\lambda)x(1-x)}
{k_{\perp}^2-x(1-x)\mu^2(\lambda)+\mu^2(\lambda)}
-\frac{\mu^2(\lambda)}
{k_{\perp}^2-x(1-x)\mu^2(\lambda)+\mu^2(\lambda)}
\right)
\nonumber\\
&\times&
\left(1+\frac{1}{x^2}+\frac{1}{(1-x)^2}
\right)
\,,\label{eq:4.1}\end{eqnarray}
where similarity function,
with momentum transfer $Q_1^2(\lambda)$ given in Eq. (\ref{eq:2.10}), 
regulates both 
UV divergences in transverse direction and IR
divergences in longitudinal direction 
\begin{eqnarray}
&& k_{\perp max}^2=
(\lambda^2+\mu^2(\lambda)) x(1-x)-\mu^2(\lambda)
\nonumber\\
&& \frac{\mu^2(\lambda)}{\lambda^2+\mu^2(\lambda)}
\leq x\leq 1 -\frac{\mu^2(\lambda)}{\lambda^2+\mu^2(\lambda)}
\,.\label{eq:4.2}\end{eqnarray}
Integration over $k_{\perp}$ gives
\begin{eqnarray}
\mu^2(\lambda) &=& \tilde{\mu}^2
+\frac{g^2C_a}{8\pi^2}\int_{x_{min}}^{x_{max}}dx
\left(
\mu^2(\lambda)x(1-x)
(1+\frac{2}{x^2})
\ln\frac{\lambda^2x(1-x)}{\mu^2(\lambda)(1-x(1-x))}\right.
\nonumber\\
&-& \left.\mu^2(\lambda)(1+\frac{2}{x^2})
\ln\frac{\lambda^2x(1-x)}{\mu^2(\lambda)(1-x(1-x))}\right.
\nonumber\\
&+&\left. (1+\frac{2}{x^2})
((\lambda^2+\mu^2(\lambda))x(1-x)-\mu^2(\lambda))
\right)
\,,\label{eq:4.3}\end{eqnarray}
where we have the symmetry with respect to interchange $x$ and $(1-x)$.
This may be simplified to
\begin{eqnarray}
\mu^2(\lambda) &=& \tilde{\mu}^2
+\frac{g^2C_a}{4\pi^2}\left(
\mu^2(\lambda)\ln\frac{\lambda^2}{\mu^2(\lambda)}
(\ln\frac{\lambda^2}{\mu^2(\lambda)}-\frac{5}{12})\right.
\nonumber\\
&+& \left.\mu^2(\lambda)(-\frac{\lambda^{2}}{\mu^{2}(\lambda)}
+\ln\frac{\lambda^{2}}{\mu^{2}(\lambda)}-\frac{5}{12})
+\lambda^{2}(-\frac{11}{12})\right.
\nonumber\\
&-& \left.\frac{1}{2}\mu^2(\lambda)\int_{x_{min}}^{x_{max}}dx(1+\frac{2}{x^2})(1-x(1-x))
\ln\left|\frac{x(1-x)}{1-x(1-x)}\right|\right)
\,.\label{eq:4.4}\end{eqnarray}
We take into account the dependence of coupling constant on
the cut-off, $g(\lambda)$, in Eq. (\ref{eq:4.4}). 
Generally, the following terms contribute to 
the right hand side of Eq. (\ref{eq:4.4})
\begin{eqnarray}
\mu^2(\lambda)&=&\tilde{\mu}^2+g^2(\lambda)
\left(\mu^2(\lambda)\ln\frac{\lambda^2}{\mu^2(\lambda)}
(c_1\ln\frac{\lambda^2}{\mu^2(\lambda)}+c_2\frac{\lambda^2}{\mu^2(\lambda)}+c_3)\right.
\nonumber\\
&+&\left.\mu^2(\lambda)(c_2^{\prime}\frac{\lambda^2}{\mu^2(\lambda)}+c_3^{\prime})
\right)
\,,\label{eq:4.5}\end{eqnarray}
where $c_i$ and $c_i^{\prime}$ are some numerical constants.
Following the same itterative procedure as outlined in the main text
we substitute the leading order value for the effective mass
into the right-hand side of Eq. (\ref{eq:4.5}), $\mu^2(\lambda)=\tilde{\mu}^2$.
The next to leading order reads
\begin{eqnarray}
\mu^2(\lambda)=\tilde{\mu}^2+g^2(\lambda)
\left(\tilde{\mu}^2 \ln\frac{\lambda^2}{\tilde{\mu}^2}
(c_1\ln\frac{\lambda^2}{\tilde{\mu}^2}+c_2\frac{\lambda^2}{\tilde{\mu}^2}+c_3)
+\tilde{\mu}^2(c_2^{\prime}\frac{\lambda^2}{\tilde{\mu}^2}+c_3^{\prime})
\right)
\,.\label{eq:4.6}\end{eqnarray}

One may consider the following scenery, how the effective gluon mass $\tilde{\mu}$
appears in the theory. In our case the effective gluon mass
plays the role of IR regulator.
It is well known, that QCD (in chiral limit, $m\rightarrow 0$) initially
does not have any scale, i.e. it is conformal invariant.
When QCD is evolved from the bare UV cut-off to some lower energy scale, 
the canonical operators are changing with renormalization group equations
in a way that the physical observables are expressed only through 
the renormalzation group invariant (cut-off independent)
combinations. From Callan-Symanzik equation the RG invariant combination  
in the leading order of perturbation theory is
\begin{eqnarray}
\Lambda=\lambda{\rm exp}\left(-\frac{8\pi^2}{b g^2(\lambda)}\right)
\,,\label{eq:4.7}\end{eqnarray}
so, $d\Lambda/d\lambda=0$. Here $\lambda$ is the running cut-off
and $\Lambda=\Lambda_{QCD}$ is the hadron scale, $\Lambda\ll\lambda$; 
$b=\frac{11}{3}N_c-\frac{2}{3}N_f$ 
is the one-loop Gell-Mann-Low coefficient in $\beta$-function.
The hadron scale, $\Lambda$, arises through
the dimensional transmutation: one introduces the renormalization point
-- the parameter with the dimension of energy, $\Lambda$, 
in order to express the running coupling constant, $g(\lambda)$,
through the coupling at renormalization point, $g(\Lambda)$, 
which is dimensionless \cite{Weinberg}.

We express all operators with dimension of energy through
the hadron scale, since it is the only scale provided after renormalization.
Therefore 
\begin{eqnarray}
\tilde{\mu}^2\sim \Lambda^2=\lambda^2 
{\rm exp}\left(-\frac{8\pi^2}{b g^2(\lambda)}\right)
\,,\label{eq:4.8}\end{eqnarray}
where in order to insure the boost invariance 
the cut-off is rescaled $\lambda\rightarrow \lambda^2/q^+$,
and the same for $\Lambda$.
The logarithmic term in Eq. (\ref{eq:4.6}) is given
\begin{eqnarray}
g^2(\lambda)\ln\frac{\lambda^2}{\tilde{\mu}^2}=\frac{8\pi^2}{b}
\,,\label{eq:4.9}\end{eqnarray}
that reduces the Eq. (\ref{eq:4.6}) to
\begin{eqnarray}
\mu^2(\lambda) = a_1 \tilde{\mu}^2\ln\frac{\lambda^2}{\tilde{\mu}^2}
+a_2\tilde{\mu}^2 + a_3 \lambda^2 + O(g^2(\lambda))
\,,\label{eq:4.10}\end{eqnarray}
where $a_i$ are numerical constants.
To remove the leading cut-off dependence 
when $\lambda=\Lambda\rightarrow\infty $, we renormalize
this equation by adding the mass counterterm
\begin{eqnarray}
m_{CT}^2=-a_3\Lambda^2
\,.\label{eq:4.11}\end{eqnarray}
The nonperturbative effective mass is given
\begin{eqnarray}
\mu^2_{NP}(\lambda) = a_1 \tilde{\mu}^2\ln\frac{\lambda^2}{\tilde{\mu}^2}
+a_2\tilde{\mu}^2
\,.\label{eq:4.12}\end{eqnarray}
Indeed, this equation is of a nonperturbative kind, since it does not
involve powers of the coupling constant.
It was possible to get a nonperturbative result,
because we regulated IR singularity  
by the hadron scale, which is given by a nonperturbative expression,
Eq. (\ref{eq:4.7}): the argument of the exponent, $8\pi^2/g^2$, can never be obtained
in perturbation theory as an expansion in powers of coupling constant.    

In fact, we use a nonzero gluon mass $\tilde{\mu}$ as a parameter
in our calculations, and at the end we take $\tilde{\mu}\rightarrow 0$.
However, from Eq. (\ref{eq:4.12}), when a mass parameter is removed, 
$\tilde{\mu}\rightarrow 0$, an effective (nonperturbative) gluon mass
vanishes (since $\tilde{\mu}$ is the only scale available
through which $\mu^2_{NP}(\lambda)$ is expressed).
In the main text we introduce therefore an additional scale $u$
to regulate IR divergences.

\end{document}